\documentclass[sigconf]{acmart}

\usepackage{siunitx}
\usepackage{graphicx}
\usepackage{subcaption}
\usepackage{balance}
\usepackage{comment}
\usepackage{amsmath}
\usepackage{mathtools}
\usepackage{hyperref}

\usepackage{balance}

\setcopyright{acmlicensed}

\copyrightyear{2019} 
\acmYear{2019}
\acmConference[MEMSYS '19]{the International Symposium on MemorySystems}{September 30-October 3, 2019}{Washington, DC, USA}
\acmPrice{15.00}
\acmDOI{10.1145/3357526.3357553}
\acmISBN{978-1-4503-7206-0/19/09}

\usepackage{multirow}

\begin{document}
	\title{ARC: DVFS-Aware Asymmetric-Retention STT-RAM Caches for Energy-Efficient Multicore Processors}

\author{Dhruv Gajaria and Tosiron Adegbija}
\email{{dhruvgajaria,tosiron}@email.arizona.edu}
\affiliation{
       \institution{Department of Electrical \& Computer Engineering \\University of Arizona}
        \city{Tucson, AZ}
        \country{USA}}



\begin{abstract}
Relaxed retention (or volatile) spin-transfer torque RAM (STT-RAM) has been widely studied as a way to reduce STT-RAM's write energy and latency overheads. Given a relaxed retention time STT-RAM level one (L1) cache, we analyze the impacts of dynamic voltage and frequency scaling (DVFS)---a common optimization in modern processors---on STT-RAM L1 cache design. Our analysis reveals that, apart from the fact that different applications may require different retention times, the clock frequency, which is typically ignored in most STT-RAM studies, may also significantly impact applications' retention time needs. Based on our findings, we propose an asymmetric-retention core (ARC) design for multicore architectures. ARC features retention time heterogeneity to specialize STT-RAM retention times to applications' needs. We also propose a runtime prediction model to determine the best core on which to run an application, based on the applications' characteristics, their retention time requirements, and available DVFS settings. Results reveal that the proposed approach can reduce the average cache energy by 20.19\% and overall processor energy by 7.66\%, compared to a homogeneous STT-RAM cache design.
\end{abstract}




\maketitle
\section{Introduction}
Caches remain very important components of modern-day and emerging processors due to their impact on both performance and power. While caches bridge the well-known processor-memory performance gap, on-chip caches can also consume a substantial amount (over 40\%) of processors' power \cite{Montanaro199749}. Thus, caches have been the focus of much research for power/energy and performance optimization.  One such optimization involves replacing traditional SRAMs with the non-volatile spin-transfer torque RAM (STT-RAM) in emerging caches. STT-RAMs have substantially lower leakage power and higher density than SRAMs \cite{Jog12}, \cite{Dong08}, \cite{Sun11}. However, STT-RAMs have significant dynamic overheads due to long write latency and high dynamic write energy \cite{Sun11}. 
 
In the absence of a power source, STT-RAMs intrinsically preserve data for up to 10 years. This duration is called the \textit{retention time}. Prior work \cite{Smullen11} showed that the long write latency and high dynamic write energy directly result from this long retention time. As such, prior works \cite{LARS8342053}, \cite{Smullen11}, \cite{Sun11}, \cite{Jog12} have studied the benefits of reducing/relaxing the retention times. Relaxing the retention time, especially in caches, is beneficial since data blocks are usually only needed in the cache for a short duration (typically less than one second). 
 
However, due to the reduced retention time, cache blocks may sometimes expire before they are evicted (or invalidated). This premature expiry results in a type of cache misses that we refer to as \textit{expiration misses}---misses that occur because a referenced block prematurely expired due to elapsed retention. To prevent premature expiry, Dynamic Refresh-Scheme (DRS) \cite{Sun11} was proposed to refresh the cache block before the data expires. However this technique accrues additional hardware costs and needs multiple read/write operations for each refresh; this limits the optimization potential \cite{Li13}. Various techniques have been proposed \cite{Li13},\cite{Jog12} to reduce the number of refreshes in order to mitigate the overheads. These techniques use either compiler optimization techniques or physical circuits, which result in additional design, power and area overheads, especially in complex multicore systems. 

Another increasingly popular optimization for emerging caches involves combining both STT-RAM and SRAM in a \textit{hybrid architecture} to take advantage of STT-RAM's low read energy and SRAM's low write energy. For instance, Li et al. \cite{Hybrid_cache2} explored hybrid caches comprising of both SRAM and STT-RAM, wherein write-intensive cache blocks were serviced by the SRAM cache and read-intensive blocks were serviced by the STT-RAM cache. Applications were scheduled to the cache that best satisfied the applications' execution needs. Donyanavard et al. \cite{Hybrid_cache1} proposed to deploy a combination 'fast cores' using SRAM caches and 'slow cores' using STT-RAM caches for multicore processors.

In this paper, we focus on exploiting the energy benefits of reduced retention STT-RAM caches in multicore processors. We explore the interplay of variable retention times and expiration misses in level one (L1) STT-RAM caches, considering a system without any refresh mechanisms. Importantly, we perform this study within the context of variable clock frequencies, which have been ignored in prior studies of STT-RAM caches. Variable clock frequencies, otherwise known as dynamic frequency scaling (DFS) or dynamic voltage frequency scaling (DVFS) \cite{dvfs_intro_ref}, is widely used in modern processors. In DVFS-enabled systems, the clock frequency is changed according to workload requirements in order to manage the power consumption, temperature, or performance \cite{DTM1}, \cite{DTM2}.

Prior work \cite{diminshing_returs} found diminishing returns in using DVFS in modern computer systems featuring SRAM caches. However, we observed that the trends were different in systems with STT-RAM caches---DVFS still offers energy saving benefits if the interplay of the clock frequency and retention time is carefully considered. The analysis presented herein is motivated by two important observations. First, due to the reduction in clock period concomitant to frequency increases, the number of cache access cycles will increase as the frequency increases. Since STT-RAMs have a higher cache write access latency than SRAMs, we observe a higher change in cache write access cycles as the frequency increases. Secondly, for STT-RAM cache, whereas reducing the retention time will typically increase the expiration misses---a major performance indicator for STT-RAM caches---increasing the clock frequency may result in the eviction of data blocks since the frequency of cache accesses also increases, thereby mitigating the expiration misses.

These observations motivate us to perform a thorough analysis of retention times for a variety of workload characteristics within the context of DVFS. Based on our analysis, we explore the appropriate retention times for different execution scenarios and objective functions (energy and  latency). We investigate the interplay of retention time specialization (via \textit{asymmetric-retention cores}) and DVFS for energy savings in multicore STT-RAM caches.

 Our major contributions are summarized as follows:
 \begin{itemize}
     \item For the first time, to our best knowledge, we analyze the impact of DVFS on reduced retention STT-RAM caches from the perspective of the cache and the overall processor. We explore the impact of frequency on STT-RAM cache misses and how it impacts the selection of the appropriate retention times for energy savings or performance improvement.
     \item For runtime retention time specialization to applications' requirements, we propose \textit{asymmetric-retention cores (ARC)}. ARC equips cores in a multicore system with different retention times and frequency ranges, such that applications are executed on the core that best satisfies their execution needs. Furthermore, we propose a low-overhead machine learning-based prediction algorithm that maps applications to the appropriate cores during runtime. 
     \item Using extensive simulations with a total of 30 applications (for training and testing), we show that our proposed work reduces the cache energy in a multicore system by 20.19\%, on average, compared to a homogeneous STT-RAM system. Furthermore, our approach reduces the overall processor energy by 7.66\% and 13.66\%, on average, compared to a homogeneous multicore STT-RAM and SRAM system, respectively. 
     \item Finally, we augment our approach to cater to a system with different execution performance constraints, and study the impact of the different constraints on our approach. We show that our prediction algorithm also works under various performance constraints.
 \end{itemize}


\section{Background and Related Work}

To provide a background for our work, this section first presents an overview of an STT-RAM cell. Thereafter, to provide context for our work, we describe prior related work on STT-RAM caches, dynamic voltage and frequency scaling, and heterogeneous multicore architectures.

\subsection{Overview of STT-RAM Caches}
\begin{figure}[t]
      \centering
      \includegraphics[width=0.8\linewidth]{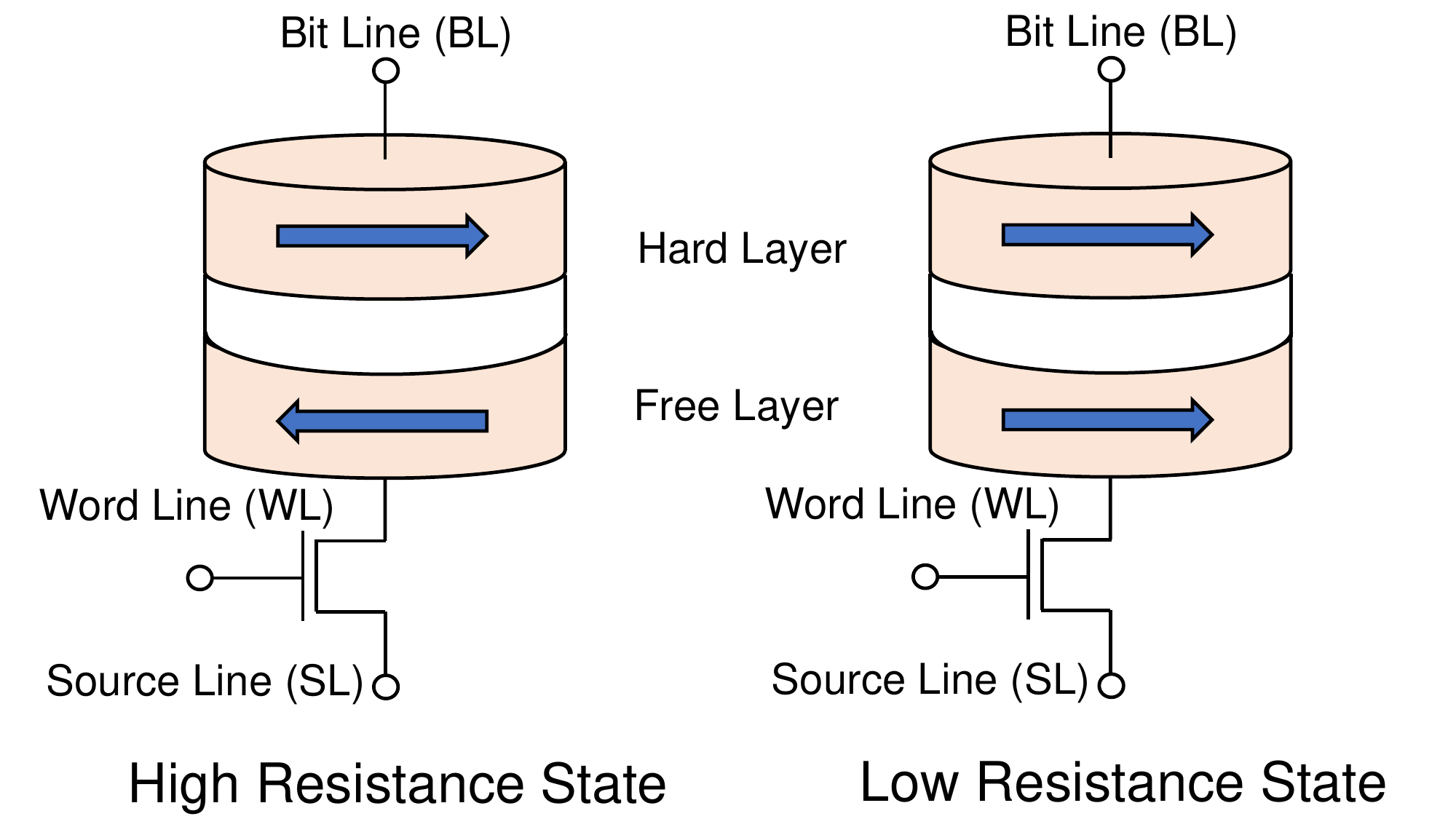}
      \vspace{-3pt}
      \caption{STT-RAM cell structure. High resistance state is in anti-parallel state and low resistance state is parallel state}
      \vspace{-5pt}
      \label{fig:cell}
\end{figure}

Figure \ref{fig:cell} illustrates the STT-RAM bit cell structure, which consists of a transistor and a magnetic tunnel junction (MTJ). In summary, MTJ comprises of two ferromagnetic layers separated by an oxide layer. The magnetization direction of the hard layer is fixed while the magnetization direction of free layer, which stores the memory bit, can be set by passing current through it. Details of the characteristics and functioning of the STT-RAM are provided in the prior work \cite{Chun13}. 

STT-RAM is a promising candidate for the cache hierarchy due to several characteristics, such as good scalability, high endurance (compared to other NVM technologies \cite{stt_adv,Smullen11}),  near-zero leakage power, and radiation hardness. STT-RAM cells require a cell area from 1/9 to 1/3 that of SRAM, which enables a larger cache with the same die footprint \cite{stt_adv}. However, deploying STT-RAM technology in caches is challenging because of high write latency and high dynamic write energy \cite{Smullen11}. Smullen et al. \cite{Smullen11} proposed relaxing the retention time of STT-RAM and observed that relaxing the retention time of STT-RAM cell reduces the latency and energy for write operations. STT-RAM can be designed with the desired retention time based on the understanding that the retention time is exponentially proportional to the magnetization stability energy, $\Delta$, which can be expressed as:

\begin{equation*}
    \Delta \propto \frac{V.H\textsubscript{k}.M\textsubscript{s}}{T}
\end{equation*}
where V is activation volume for STT-RAM writing current, H\textsubscript{k} is in-plane  anisotropy field, M\textsubscript{s} is saturation magnetization and T is absolute temperature in Kelvin \cite{Smullen11}.

We note that STT-RAM also has other issues, such as reliability and process variation issues, that still impede its widespread adoption in modern processors. While these issues are outside the scope of this paper, there is much ongoing research to address them. For instance, Emre et.al \cite{stt-ram_reliability} studied the reliability of STT-RAM cell and showed that process variations and variations in device edge geometry affects their failure rate. They used error correction codes and circuit-level parameter tuning to obtain a block failure rate (defined as a binomial distribution of uniform errors, of \num{e-9}. Similarly, Alkabani et.al \cite{reliability_2} used a combination of pulse width scaling and error correction coding schemes for STT-RAMs to reduce the write energy consumption. The proposed approach reduced the average write energy by 46\%  with uncorrectable bit error rate (UBER) of \num{e-13}.

\subsection{Mitigating the Overheads of Relaxed Retention Time STT-RAM}
Prior work \cite{Sun11} showed that relaxing the retention time of an STT-RAM cache could improve the IPC and energy by up to 80\% and 84\%, respectively, compared to non-volatile STT-RAM caches. However, relaxed retention STT-RAMs can still result in energy or time overheads resulting from premature expiration of data blocks. To mitigate the overheads of premature expiration, prior work also proposed various dynamic refresh schemes (DRS) \cite{Sun11}, \cite{Jog12}. DRS refreshes data blocks if they must remain in the cache after the retention time elapses. These approaches typically use counters to monitor the block's lifetime in order to determine when to refresh the block. However, these techniques incur additional overheads, resulting from the refresh buffers and refresh operations. As such, optimization potential may be severely limited and other studies have been done to explore techniques for reducing the cost of refresh operations \cite{Li13}.

Kuan et al. \cite{LARS8342053} found that different applications require different retention times at a finer granularity than previously shown. They focused on mapping different applications with their best retention times, leading to latency and energy savings of up to 13.2\% and 35.8\%, respectively, compared to DRS. However, the proposed approach required four STT-RAM units in each core, resulting in hardware overhead, which our work mitigates. Furthermore, the impacts of variable clock frequency (and frequency scaling) on applications' retention time needs have not been explored by prior studies.

\subsection{Dynamic Voltage Frequency Scaling}
Dynamic voltage and frequency scaling (DVFS) is a popular optimization technique that is implemented in several mainstream processors. DVFS adjusts the CPU voltage and frequency according to runtime task requirements. There have been several studies on the impact on DVFS on system performance and energy \cite{dfs_4,dfs_3,dvfs_wrokload}; we focus herein on the impact of DVFS on the cache.  

Prior works have studied the impacts of voltage and frequency scaling on SRAM architectures \cite{sram_dfs1,sram_dfs2}. For instance, Wang et al. \cite{dcr_dvs2} used a combination of dynamic cache reconfiguration and DVS to achieve energy savings compared to a DVS-only system. To reduce energy, Saito et al. \cite{diff_sram_config_dvfs} suggested switching between high and low speed SRAM L1 caches operating at variable frequencies according to the requirements of executing applications. While these previous works used variable cache sizes, in our work, we keep the cache size constant and focus on the STT-RAM cache's retention time variability. However, we note that the idea of variable cache sizes is orthogonal to the work presented herein, and we leave this exploration for future research.  

Peneau et al. \cite{LIRMM} discussed the impact of frequency on cache access cycles for SRAM and STT-RAM. However, the authors focused their analysis on the selection of operating frequencies that have same latencies for SRAM and STT-RAMs for loop nest optimization. To the best of our knowledge, no prior study has analyzed how retention time design choices are affected by variable clock frequencies. Thus, we focus our analysis on the impact of variable frequencies on the right choice of retention times for executing applications' requirements, and suggest designs that leverage the synergy of DVFS and various retention times for energy savings.

\subsection{Heterogeneous Multicore Architectures}

There has been much prior work on heterogeneous multicore architectures to enable the specialization of system resources to application requirements. These prior techniques have been leveraged for energy reduction, load balancing, thermal management, etc. \cite{het_arch1, het_arch2, het_arch3, het_arch4}. In addition, there is much prior work on scheduling of applications to the appropriate core during runtime \cite{het_sched1, het_sched2, het_Sched3}. Even though several system components (e.g., branch predictors, in-order vs out-of-order execution hardware, etc.) can be leveraged for heterogeneity, we focus on using the benefits of variable frequencies in synergy with heterogeneous retention time STT-RAM cores, which we call \textit{asymmetric-retention cores (ARC)}, for energy savings in a multicore system. We believe that the work proposed herein will be beneficial for current and emerging heterogeneous core systems, such as the big.LITTLE cores \cite{big_little}.


\section{Interplay of DVFS and STT-RAM Caches} \label{sec:interplay}

In the analysis presented herein, we focus on L1 data cache since it exhibited more application variability than the instruction cache. A carefully chosen static instruction cache retention time---we empirically selected a 100$ms$ retention time---sufficed for the considered applications in our work (similarly to prior work \cite{LARS8342053}). In this section, we analyze the impact of clock frequency variations on STT-RAM cache retention time requirements and expiration misses to provide a basis for the proposed ARC architecture.

\subsection{{Impact of Variable Clock Frequency on STT-RAM Caches}\label{freq_behaviour}}

\begin{figure}[t]
\centering
    \begin{subfigure}[t]{\linewidth}
      \centering
      \includegraphics[width=0.8\linewidth]{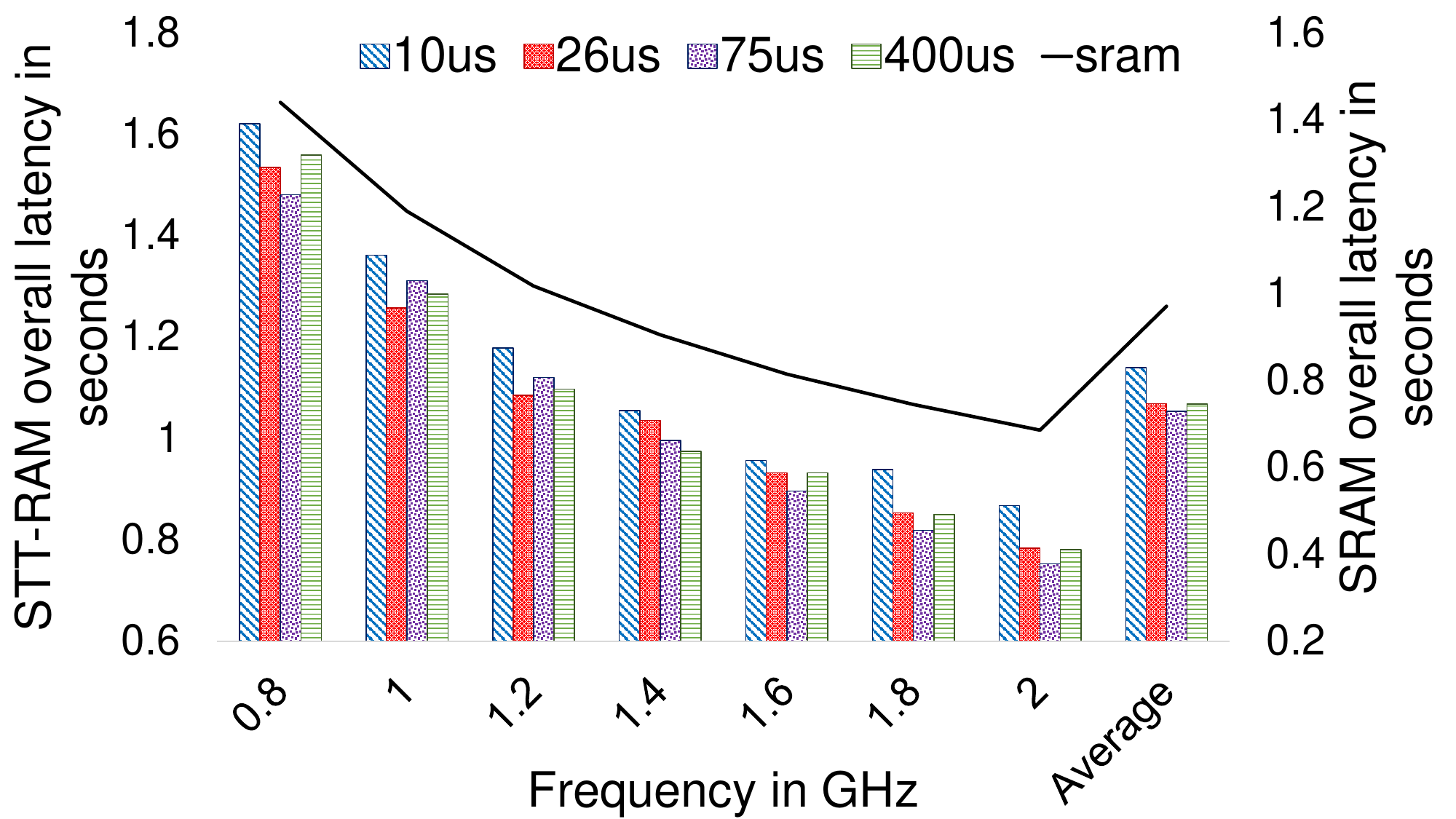}
      \vspace{-3pt}
      \caption{Latency}
      \label{fig:freq_sim_sec}
    \end{subfigure}%

    \begin{subfigure}[t]{\linewidth}
      \centering
      \includegraphics[width=0.8\linewidth]{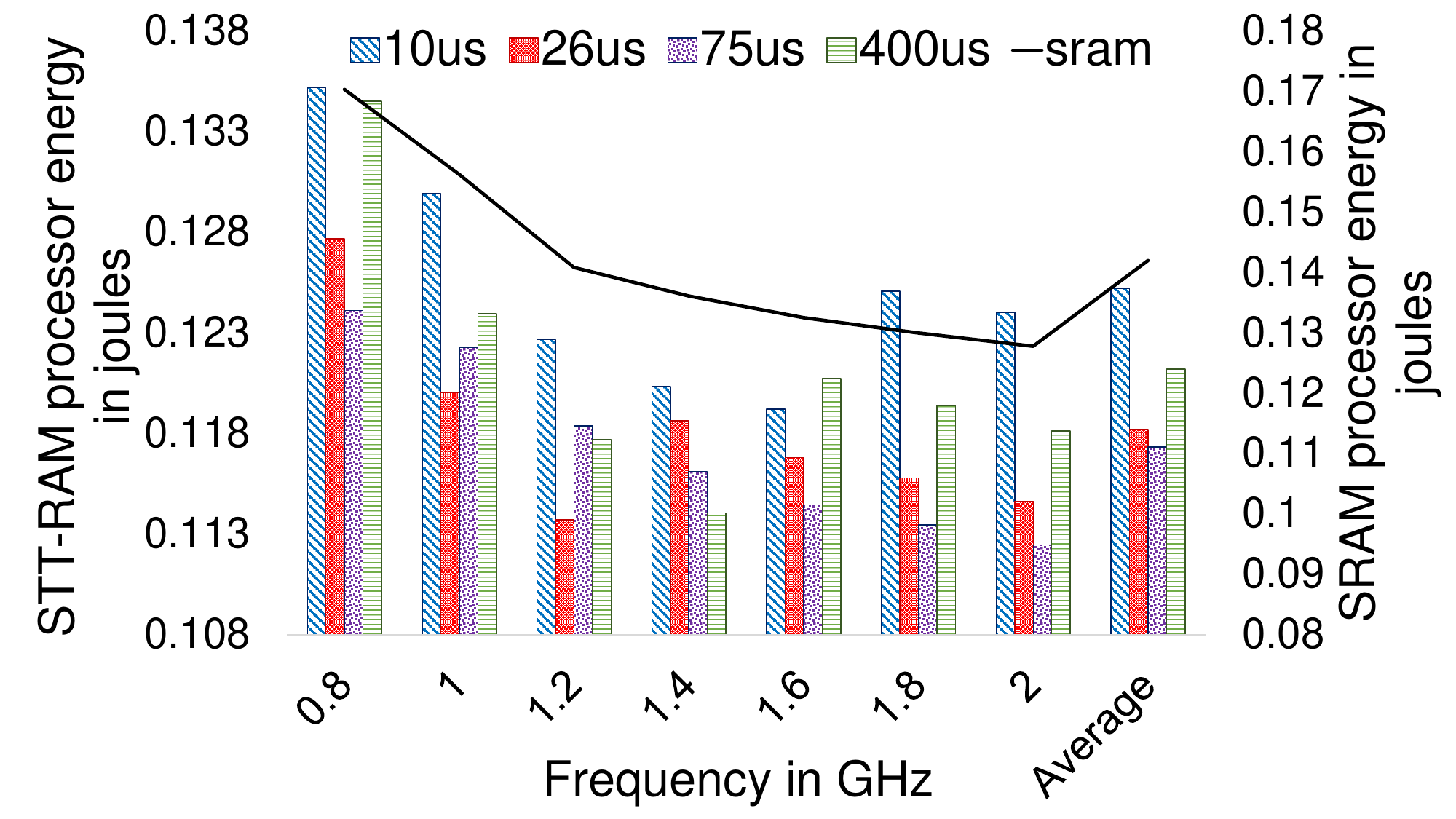}
      \vspace{-3pt}
      \caption{Processor Energy}
      \vspace{-5pt}
      \label{fig:freq_proc}
    \end{subfigure}
\caption{Impact of frequency scaling to performance and processor energy compared to SRAMs. SRAM caches are faster than STT-RAM caches but consume high energy}
\label{fig:feature_selection}
\vspace{-15pt}
\end{figure}

To explore the impact of variable clock frequency on STT-RAM caches in comparison to SRAM caches, we performed experiments using 30 benchmarks from the SPEC 2006 \cite{spec2006}, MiBench \cite{mibench}, and GAP \cite{beamer2015gap} benchmark suites to represent a variety of applications. An application's retention time requirement is a function of the application's \textit{cache block lifetimes}---the duration for which a block must remain in the cache before it is evicted or invalidated \cite{Jog12}. Thus, to empirically determine a set of retention times for our experiments, we exhaustively analyzed the average cache block lifetimes of our benchmarks to determine the range of retention times that satisfied the different benchmarks' needs. We considered several retention times, but, based on the applications' characteristics, narrowed our focus down to: 10$\mu$s, 26.5$\mu$s, 75$\mu$s, and 400$\mu$s. (details of our experimental setup are in Section \ref{sec:setup}). 

We observed and analyzed the changes in the cache and processor energy and latency with different retention time and frequency values. Ideally, the goal is to keep the cache access cycles (read and write) as low as possible. However, as expected, the clock frequency dictates the changes in the cache access cycles. The access cycles can be expressed with respect to frequency and access latency as $access\_cycles = ceil (frequency * access\_latency)$. Whereas the read and write latencies remain constant for SRAM, for STT-RAM, the read latency changes very slightly across different retention times, while the write latency increases more drastically as the retention time increases. As such, we observe a more pronounced impact of frequencies on STT-RAM caches than on SRAM caches. 

Figure \ref{fig:freq_sim_sec} depicts the latency achieved using the STT-RAM (with different retention times) and SRAM caches. All other processor configurations remain constant. Each bar and line represents average results across all the benchmarks.  Our first observation from the figure is that the best frequency for SRAM, with respect to latency, is the highest---2GHz. For the STT-RAM system, on the other hand, the best retention time for latency varies with different frequencies---longer retention times do not necessarily improve the performance compared to shorter retention times. For instance, at 0.8GHz frequency, 75$\mu$s had the shortest latency as it requires one cycle per write operation, whereas at 1GHz to 1.2 GHz, 26$\mu$s performed better for latency due to the lower write cycles than 75$\mu$s and 400$\mu$s. We also observed that apart from the access cycles, the \textit{expiration misses} also impacted the retention time requirements. For example, although 10$\mu$s and 26.5$\mu$s have equal cache access cycles, we observed that benchmarks exhibited fewer expiration misses on the 26.5$\mu$s cache, since several more cache blocks were prematurely evicted on the 10$\mu$s cache. As such, the 26.5$\mu$s cache was more favorable for latency than 10$\mu$s. Additional analysis of expiration misses is presented in Section \ref{section:exp_miss}. 

 
Figure \ref{fig:freq_proc} depicts the changes in overall processor energy achieved by the STT-RAM and SRAM caches for different frequencies. Much like the latency results, the best frequency for energy in the SRAM cache is 2GHz. Even though the 2GHz would have a higher power consumption than lower frequencies, benchmarks are run much faster at 2GHz resulting in overall lower energy. These observations are similar to prior work \cite{diminshing_returs}. However, we observed that the best frequency for energy changed for different retention times---higher frequencies were not necessarily always better. For instance, for 10$\mu$s, the best frequency is 1.6GHz, for 26.5$\mu$s the best is 1.2GHz, for 75$\mu$s the best is 2.0GHz, and for 400$\mu$s the best is 1.4GHz. Note that these results are on average across all the benchmarks; we observed variability among the benchmarks, revealing optimization potential when DVFS is employed in synergy with different retention times. These variable behaviors, similar to the latency results, result from the variance in cache access cycles and expiration misses, which we discuss in the following subsection.

\subsection{{Impact of Frequency on Expiration Misses}\label{section:exp_miss}}

\begin{figure}[t]
      \centering
      \includegraphics[width=1\linewidth]{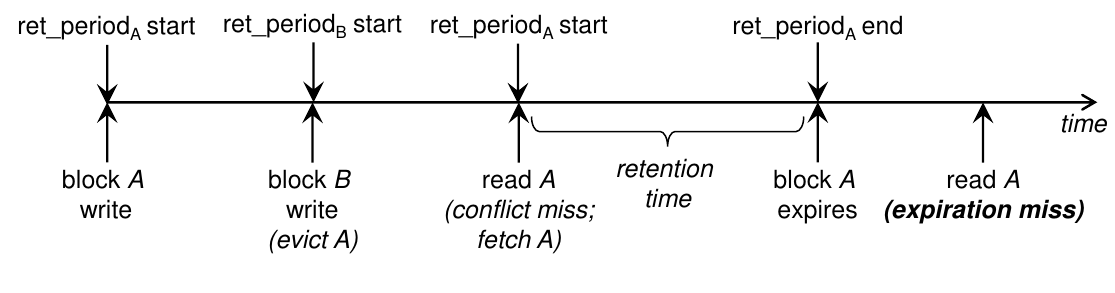}
      \vspace{-3pt}
      \caption{Illustration of expiration misses. Assume that blocks A and B are in the same memory location, i.e., a write from one block would evict the currently resident block}
      \vspace{-5pt}
      \label{fig:expirationMisses}
\end{figure}

\begin{figure}[t]
      \centering
      \includegraphics[width=0.8\linewidth]{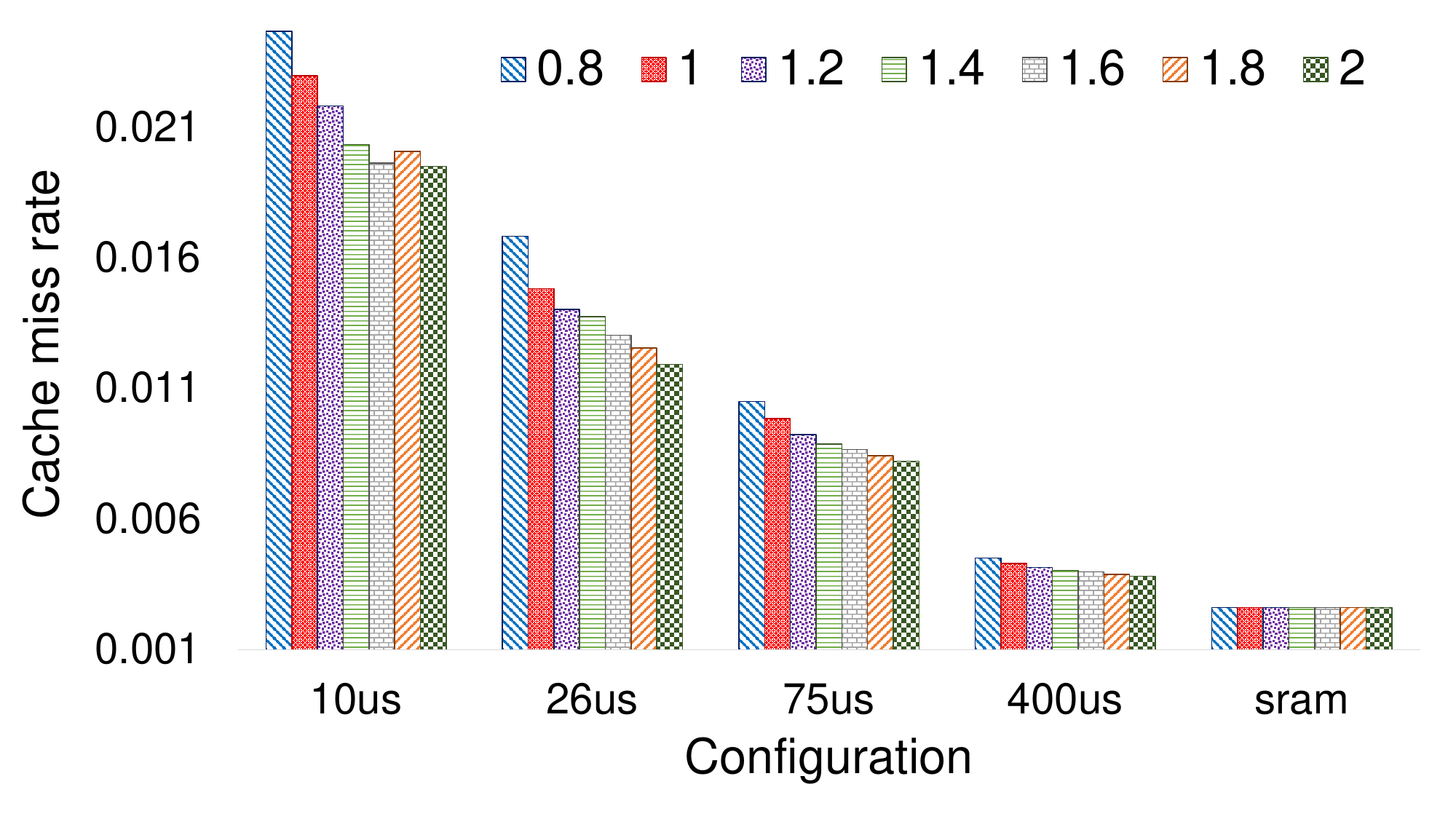}
      \vspace{-3pt}
      \caption{Change in miss rate with respect to frequency. The change in miss rates is observed due to decrease in expiration misses with increase in frequency }
      \vspace{-5pt}
      \label{fig:freq_exp+_misses}
\end{figure}

\begin{figure*}[h]
      \centering
      \includegraphics[width=0.6\linewidth]{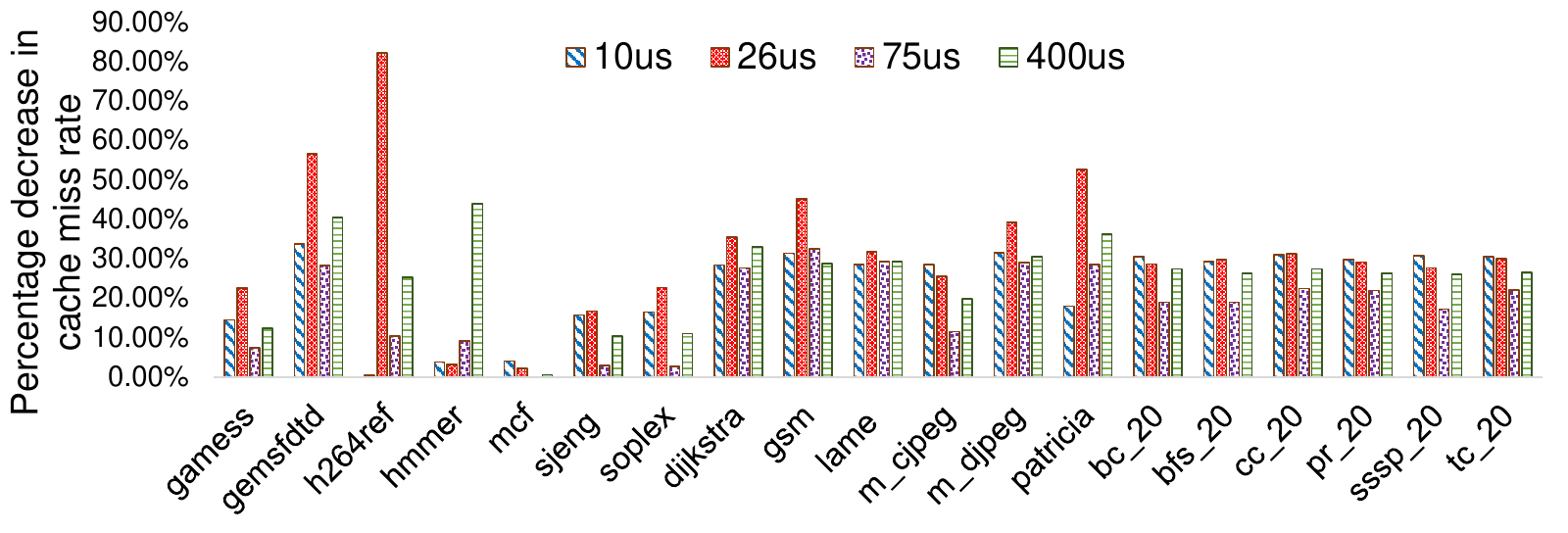}
      \vspace{-3pt}
      \caption{Decrease in cache miss rate with increase in frequency from 0.8GHz to 2.0GHz for various retention times. We observe specific retention times having high change in cache miss rates with respect to frequency due to variance in cache block lifetimes for different benchmarks}
      \vspace{-5pt}
      \label{fig:bench_exp+_misses}
\end{figure*}

We define an \textit{expiration miss} as a miss that results from a reference to a block that expired prematurely due to a short retention time. Figure \ref{fig:expirationMisses} illustrates the occurrence of an expiration miss. Assume that blocks $A$ and $B$ reside in the same memory location; that is, if $A$ is first written into the cache, a write of $B$ evicts $A$ from the cache. Thereafter, a read of $A$ results in a conflict miss; $A$ is brought into the cache, and its retention period is restarted. After the retention time elapses (at $ret\_time_A$ $end$), a reference to $A$ results in an expiration miss, since the miss only occurred because of an elapsed retention time. Therefore, shorter retention times increase the expiration misses, and in effect, overall miss rates.

We observed that due to the presence of expiration misses in reduced retention STT-RAM caches, frequency has a much larger impact on STT-RAM cache miss rates than SRAM. Figure \ref{fig:freq_exp+_misses} shows the average miss rates of our benchmarks for the different retention times and SRAM with different frequencies. Whereas the miss rates of SRAM are relatively unaffected by frequency changes, the cache miss rates vary for STT-RAMs: in general, as the frequency increases, the miss rate decreases. This is because as the frequency increases, the cache block accesses also increase. As such, blocks that may have expired at lower frequencies may be replaced due to faster computations, thereby reducing the expiration misses. For most of the retention times, the highest decrease in cache expiration misses occurs when the frequency is increased from 0.8 GHz to 1GHz. However,  due to increased cache access cycles with increase in frequency, we observe an increase in expiration misses for some cases.

To further illustrate that the impact of frequency in STT-RAM caches is also application-dependent, Figure \ref{fig:bench_exp+_misses} shows the reduction in cache miss rates observed when the frequency is increased from 0.8GHz to 2.0GHz. For brevity, the figure only shows an arbitrary subset of the considered benchmarks. The key takeaway from the figure is that the frequency-retention time interplay varies substantially for different applications. For instance, although 10$\mu$s typically has the highest miss rates, the frequency change had a higher impact for 26$\mu$s for some benchmarks, like $h264ref$, $soplex$ (from SPEC), and $patricia$ (from MiBench), whereas the highest impact observed for $hmmer$ (from SPEC) was with the 400$\mu$s cache. Apart from showing the impact of frequency and retention times on STT-RAM cache performance, these results also illustrate the variability of this interplay for different applications. This behavior makes it challenging---and necessary---to not only specialize retention times to applications requirements, but also to predict the best retention time, while considering the frequency and application-specific cache block lifetimes.

\section{DVFS-Aware Asymmetric-Retention Core (ARC)}
\begin{figure}[t]
      \centering
      \includegraphics[width=0.85\linewidth]{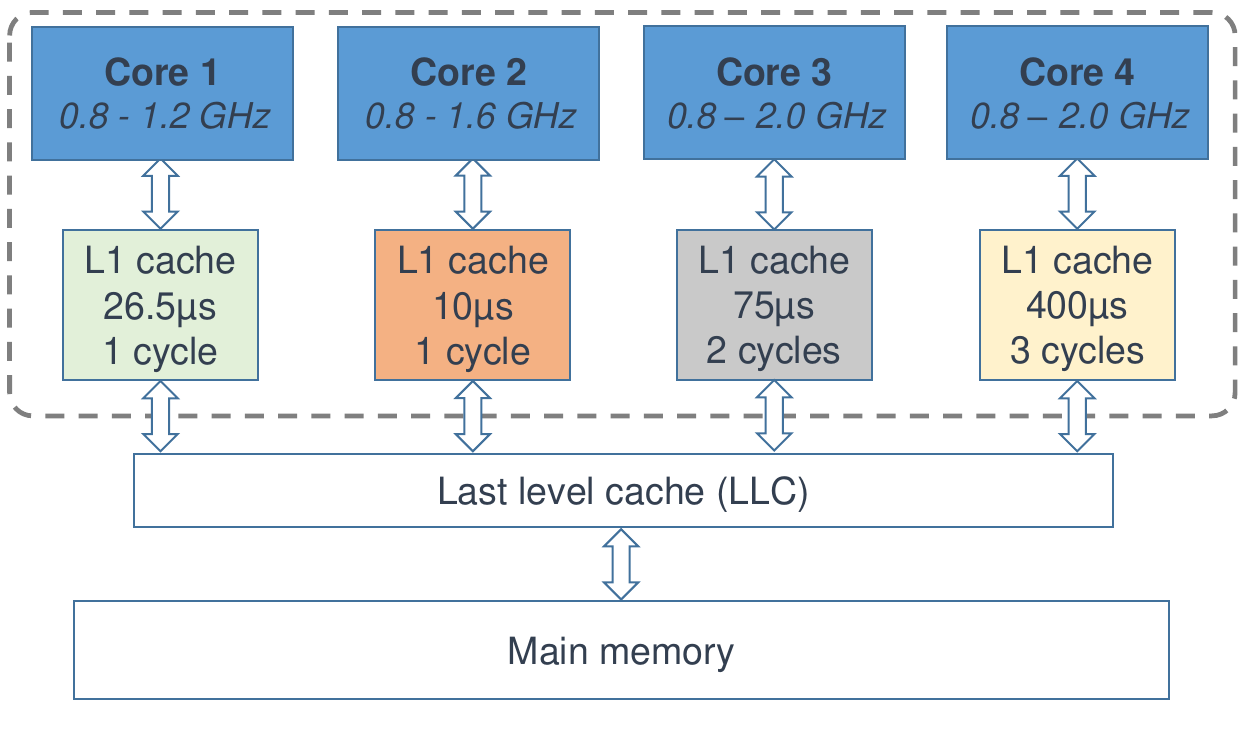}
      \vspace{-3pt}
      \caption{Proposed DVFS-aware asymmetric-retention core architecture. Each core features different frequency ranges and different retention times in the L1 STT-RAM cache, resulting in different cache write cycles at each core}
      \vspace{-5pt}
      \label{fig:arch}
\end{figure}

This section describes our proposed architecture for leveraging the interplay of DVFS and application-specific retention times for energy savings. Given the analysis in Section \ref{sec:interplay}, it is clear that a balance must be achieved between leveraging the impacts of variable retention time and clock frequency requirements, while considering these factors' impacts on expiration misses. 

\subsection{Proposed ARC Architecture}{\label{base_arc}}
We propose a DVFS-aware asymmetric-retention core (ARC) architecture featuring different retention times and frequency ranges to satisfy a variety of runtime execution needs. The ARC architecture also features a runtime prediction model that allows different applications to be scheduled to the appropriate core during runtime, based on the applications' characteristics. Note that the asymmetry of the proposed architecture can be extended to any number of configurations. For this work, however, we limit the architecture to heterogeneous L1 cache retention times and their frequency ranges, which dictate the caches' write access cycles.

Figure \ref{fig:arch} illustrates the proposed architecture, which comprises of $n$ cores ($n = 4$ in this work), each featuring a different range of clock frequencies and retention times. The range of frequencies and retention times were empirically selected, based on extensive design-time analysis, to satisfy the variety of application characteristics in our benchmarks. The configurations for a different set of applications may be different. For all the cores, the frequencies range from 0.8GHz to different caps in order to constrain the write cycles for each core. The frequency for the 10$\mu$s core is capped at 1.6 GHz and the 26.5$\mu$s core is capped at 1.2 GHz to maintain a 1-cycle cache write latency. The 75$\mu$s and 400$\mu$s cores can operate at up to 2.0 GHz, with two and three cache write cycles, respectively. 

Given the analysis in Section \ref{sec:interplay} and the characteristics of the cores in Figure 6, we expect that when performance optimization is the objective, application execution will be biased towards cores 3 and 4 (at 2 GHz). When energy optimization is the objective, there will be a higher variability in the best core. The experiments bear these out and are detailed in Section \ref{sec:evaluation}. We also observed a trend in the best frequencies for different retention times when energy is the objective function---higher frequencies were generally better for energy optimization. For instance, in general, the best frequencies for energy optimization for the 10$\mu$s and 26.5$\mu$s retention times were 1.6 GHz and 1.2 GHz, respectively, and for 75$\mu$s and 400$\mu$s, 2 GHz. Note that this is not necessarily \textit{always} the case; these were general observations. These observations play into the experiments presented herein, since our focus is on energy consumption. However, we concede that the analysis may change if other optimizations (e.g, thermal management \cite{thermal_management}), which are outside the scope of this paper, come into play. 

With reduced retention times, there is a chance for data blocks to become unstable if the retention time elapses prior to a block's eviction \cite{Sun11}. To prevent occurrences of data corruption and maintain the data integrity, the cache blocks are augmented with monitor counter bits. The counter is implemented using a $k$-state finite state machine (FSM), which is controlled by the cache controller, and has a clock period that is $k$ times smaller than the retention time. When the counter reaches state $k - 1$---i.e., before the retention time elapses---the block is first written back to lower level memory (if dirty) and invalidated. For low-overhead implementation, we assumed 2-bit counters per block in our work for a 4-state counter FSM. A similar technique has been used in prior work \cite{Jog12}.

\subsection{Overview of ARC Prediction Approach} \label{sec:approach}
\begin{figure}[t]
      \centering
      \includegraphics[width=0.6\linewidth]{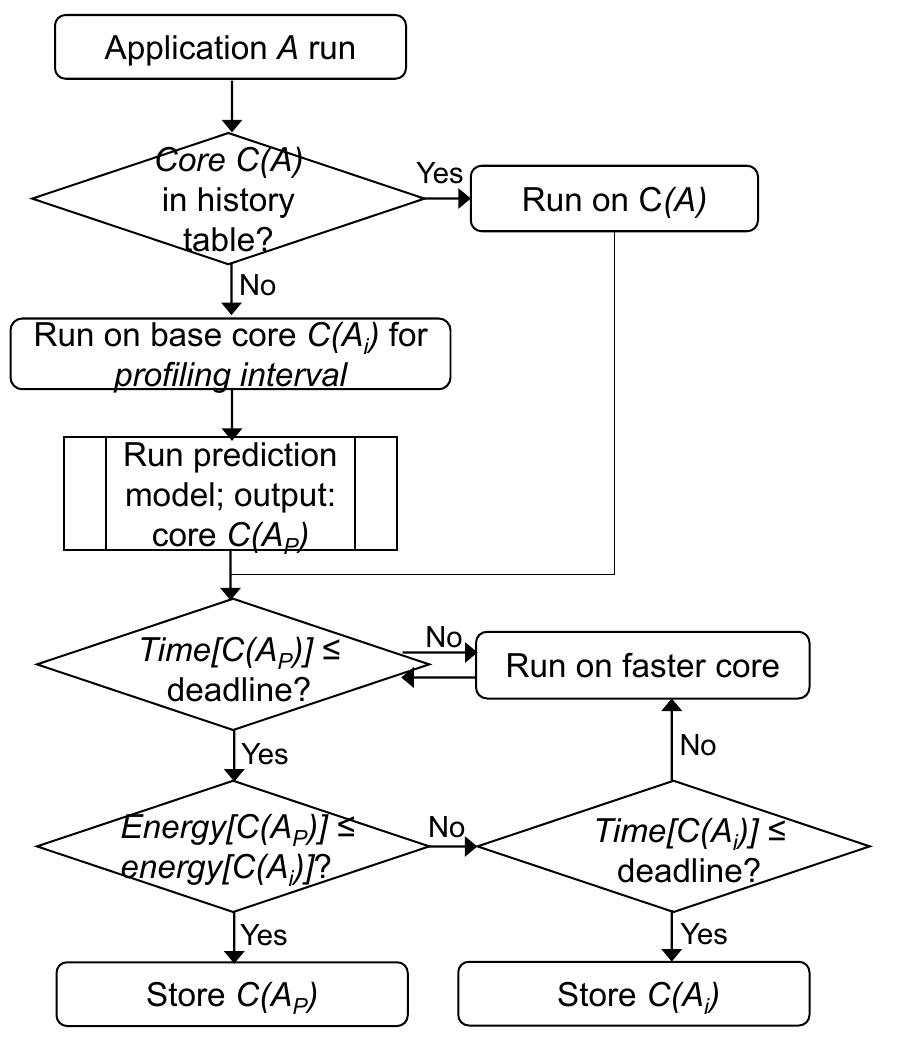}
      \vspace{-3pt}
      \caption{High-level overview of our approach. The flowchart also illustrates how runtime deadline constraints are handled}
      \vspace{-5pt}
      \label{fig:model}
\end{figure}

There are several prior works that employ machine learning for predicting the best core or configurations in computer systems \cite{multicore_ml1, multicore_ml2, multicore_ml3}, and our work is along the lines of these prior works. Our goal was to develop a low-overhead model for predicting the best core on which to run an application, based on the applications' execution characteristics. Since this is a runtime model, we also sought to develop a model that allows prediction via easily obtainable execution statistics from hardware performance counters. These requirements preclude the direct use of expiration misses  (Section \ref{section:exp_miss}) for prediction, since expiration misses are difficult to estimate during runtime. Thus, we opted to explore a combination of application characteristics that correlate with the expiration misses to enable runtime prediction of the best core for new applications. 


Figure \ref{fig:model} depicts the high-level flowchart of our approach for determining the best core on which to execute an application. When an application $A$ is run, a history table is checked to see if $A$ has been encountered before, in which case, $A$ is run on the previously stored core $C(A)$. Otherwise, $A$ is run on a \textit{base} core $C(A_i)$---the fastest available core---for a \textit{profiling interval}, which represents a brief period of time during which the application execution statistics are profiled. We experimented with various profiling intervals ranging from 1M instructions to 100M instructions, and found that 3M instructions was sufficient to obtain stable statistics for the required prediction. The statistics, including the cache statistics (e.g., miss rates, average accesses), memory controller statistics (e.g., bus write requests, bus utilization), are used as input parameters to the prediction model. Based on extensive empirical analysis, we determined that these statistics correlate strongly with an application's average cache block lifetimes and expiration miss behavior, which dictate the retention time requirements. 

If some a priori performance constraints (e.g., deadlines) are specified for the application, the constraint information can be used to select a different base core in order to improve optimization potential. For instance, if an application's worst-case execution time on the fastest core is known, and a specified timing/deadline constraint substantially relaxes the worst-case time, a slower core (e.g., Core 2 vs. Core 4 in Figure \ref{fig:arch}) can be used as the base to reduce the energy consumption of the profiling stage. Note that the worst-case execution time (WCET) analysis \cite{wcet_inst_count} is outside the scope of this paper. 

Next, the prediction model (Section \ref{section:deadline_aware_ped}) is run using statistics from the profiling stage as input, and the model outputs the best core $C(A_P)$ for running application $A$. Our approach checks that any specified deadline is met (default to $\infty$ if none is specified), and thereafter checks that the energy consumed on the predicted core $C(A_P)$ is less than the base. If the predicted core does not meet the deadline, a faster core is used (i.e., a core with a higher frequency cap), and the checking process is repeated for both time and energy. This checking process acts as a feedback that enables the best core for an application to be updated, when necessary (e.g., if that application is executed with a new input). Once a core is found that reduces the energy consumption or meets a specified deadline, the core is stored in the history table for subsequent executions of the application.

\subsection{{ARC Prediction Model}\label{section:deadline_aware_ped}}
\begin{figure}[t]
      \centering
      \includegraphics[width=0.75\linewidth]{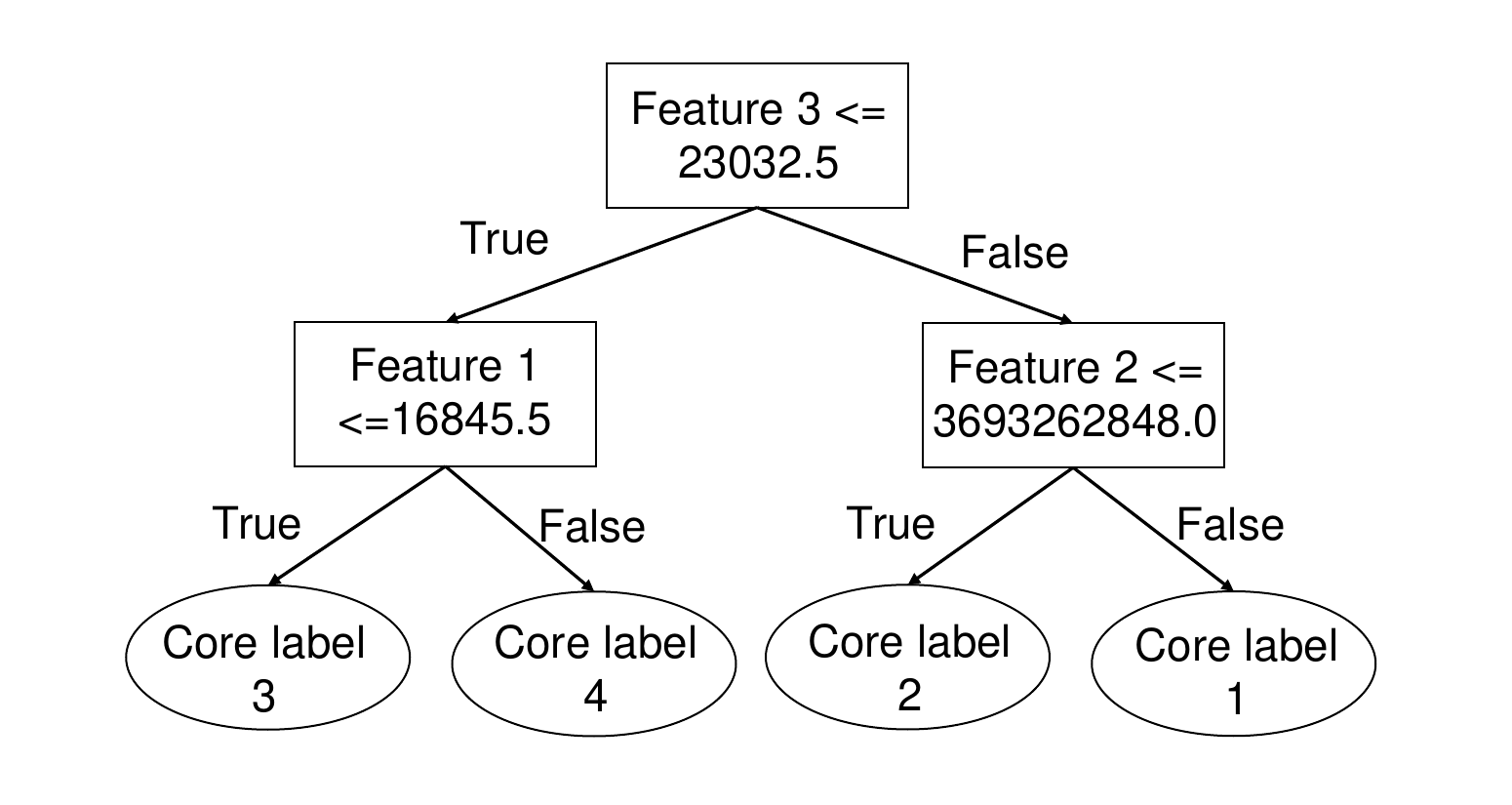}
      \vspace{-3pt}
      \caption{Illustration of decision tree for core predictions}
      \vspace{-5pt}
      \label{fig:decision_tree}
\end{figure}
\begin{figure}[t]
      \centering
      \includegraphics[width=1.0\linewidth]{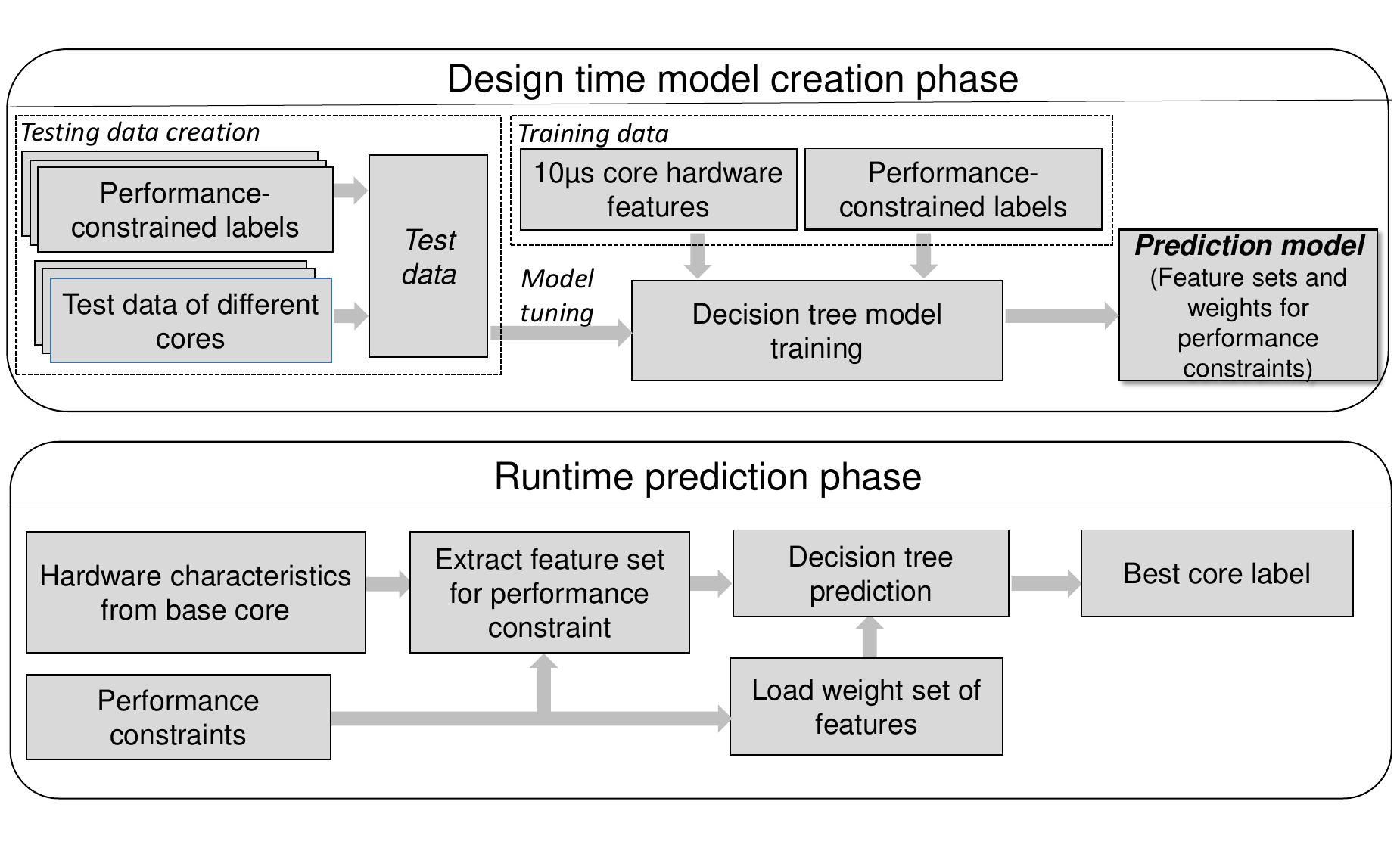}
      \vspace{-3pt}
      \caption{Flow of ARC prediction model. The model comprises of the design time model creation phase (featuring model training and testing) and runtime prediction phase}
      \vspace{-5pt}
      \label{fig:prediction}
\end{figure}

\begin{table*}[t]
\renewcommand{\arraystretch}{0.65}
\caption{Feature sets for different deadline constraints}
\label{tab:feature_set}
\centering
\vspace{-5pt}
\scalebox{0.8}{
\begin{tabular}{|c|c|c|}
    \hline
    {No constraint (best energy)} &{Best performance/10\% slack} &{20\% slack}\\  
    \hline
    L1 data cache hits &L1 data cache read misses &L1 data cache hits\\
    \hline
    L1 data cache read misses &Memory idle time &L1 data cache read accesses\\
    \hline
    L1 data cache total misses &Memory read hits &L1 data cache read misses\\
    \hline
    L1 instruction cache total misses &--&Memory idle time\\
    \hline
    Memory bus utilization for read operations &--&Memory bus utilization  for write operations\\
    \hline

\end{tabular}}
\vspace{-2pt}

\end{table*}

We explored several machine learning prediction models (e.g. Support Vector Classifiers, K-nearest neighbours, Naive Bayes, Random Forest classifier, etc.) to predict the best core/retention time, and found \textit{decision trees} \cite{decision_tree} to be the best fit for our purposes. Decision trees offers several advantages, including simplicity, low computational requirements, and fast prediction time. These features are especially important for a runtime implementation in a resource and timing constrained environment \cite{low_dataset2}. Furthermore, decision trees enable accurate predictions in the presence of small datasets, which enables us to further reduce the overhead of the prediction process \cite{low_dataset1, low_dataset2}. 

A decision tree splits a dataset into smaller subsets and uses a tree-like graph of decisions and their possible outcomes. Figure \ref{fig:decision_tree} illustrates a sample decision tree structure in our ARC prediction model. The topmost node in the tree is called the root node which is used for all the predictions. The decision tree then splits itself into either decision nodes or the leaf nodes (i.e, data output) depending on the dataset. To determine the quality of data splitting, our model uses an impurity criterion called \textit{Gini index} \cite{gini_index}, which measures how good a split is by determining how mixed the classes are in the two groups created by the split. The splitting of data and the formation of decision tree structure happen during the training stage. During the prediction stage at runtime, the corresponding hardware characteristics are input to the decision tree to predict an output core label. Depending on the data pattern, the complexity and depth of the decision tree will vary substantially. Thus, for different performance constraints, we used different tree structures. Additional low-level details and benefits of decision trees for resource-constrained prediction has been detailed in prior work \cite{resource_eff_dec_tree}.

Figure \ref{fig:prediction} presents the flow of the ARC prediction model\footnote{All our data can be found at: \url{www.ece.arizona.edu/tosiron/downloads.php}}. The model comprises of two phases: 1) the \textit{model creation phase}, which comprises of the model training and testing, and takes place at design time and 2) the \textit{prediction phase}, which occurs during runtime for unknown applications or known applications with different timing/deadline constraints. In the model creation phase, we determined the feature set and  their respective weights for different \textit{performance constraints}. The performance constraints, which can be dictated by a user-specified deadline, for example, allow the model to explicitly relax the strictness of performance optimization in order to allow for higher energy savings. We observed that different performance constraints required different models for accurate runtime prediction, due to the variation of labels for different constraints. Thus, the model creation phase outputs different models for different levels of slack from the best performance possible.

To maintain low memory footprint and complexity, we used the 10$\mu$s core (Core 2) to obtain the hardware features for the model training. We used this core because it exhibited the highest rate of expiration misses and the highest amount of variability in execution characteristics compared to the other cores. The high variability in the data ensures more exact boundary conditions, thereby enabling more accurate training of our models. To validate this choice, we also experimented with using the other cores to obtain the training data and found that the highest runtime prediction accuracy was obtained using the 10$\mu$s core to obtain the training data.

To create/train the model, we began with 41 features (execution characteristics). Using feature selection \cite{feature_selection}, we determined the most relevant features for different performance constraints, and substantially reduced the feature set to 3 to 6 depending on the performance constraints. Table \ref{tab:feature_set} lists the different performance constraints considered and their feature sets. We considered prediction scenarios for different target constraints, including no constraint (i.e., best energy), best performance possible, 10\% slack on the best performance, and 20\% slack. For the initial model creation, we used SPEC benchmarks as our training dataset, due to the diversity of execution characteristics, and used MiBench and GAP benchmarks for testing. We used a pre-trained model in order to substantially reduce runtime overheads and complexities.

During the prediction phase, the model first checks the performance constraint and then loads the respective model trained for the specified constraint. The downside of this technique is that in its current state, our approach only satisfies a limited set of performance constraints (Table \ref{tab:feature_set}). However, this does not degrade the overall effectiveness of our approach, and it can be extended for additional constraints. The corresponding features from the applications are extracted from hardware performance counters after executing them for a profiling interval. Thereafter, the appropriate weights obtained from the training phase are then used in the decision tree model \cite{decision_tree} to predict the best core label for the executing application.

\begin{table}[t]

\renewcommand{\arraystretch}{0.65}
\caption{Processor and cache configurations}
\label{tab:retention}
\centering
\vspace{-5pt}
\scalebox{0.8}{
\begin{tabular}{|c||c|cccc|}
    \hline
    Processor configuration			&\multicolumn{5}{c|}{8GB RAM, in-order, 2-wide, 0.8 -- 2 GHz}\\
    \hline
    Cache				&\multicolumn{5}{c|}{22nm, 32KB, 64B line size, 4-way}\\
    \hline
    Memory device				&SRAM	    &\multicolumn{4}{c|}{STT-RAM} \\
    \hline
    Retention times &--   &10$\mu$s     &26.5$\mu$s     &75$\mu$s   &400$\mu$s	\\
    \hline
    Hit latency      	&0.453ns	        			        &0.464ns		        &0.454ns 		        &0.445ns    &0.443ns\\
    \hline
    Write latency     	&0.312ns	        			        &0.601ns		        &0.769ns		        &0.981ns    &1.389ns \\
    \hline
    Read energy (per access)	&0.007nJ          &0.003nJ        &0.003nJ        &0.003nJ    &0.003nJ\\
    \hline
    Write energy (per access) 	&0.006nJ    &0.026nJ   &0.030nJ	            &0.035nJ        &0.045nJ    \\
    \hline
    Leakage power               &50.328mW     &\multicolumn{4}{c|}{13.1448mW}		\\

    \hline
\end{tabular}}
\vspace{-2pt}

\end{table}

\subsection{Scalability}

\begin{figure}[t]
      \centering
      \includegraphics[width=1.0\linewidth]{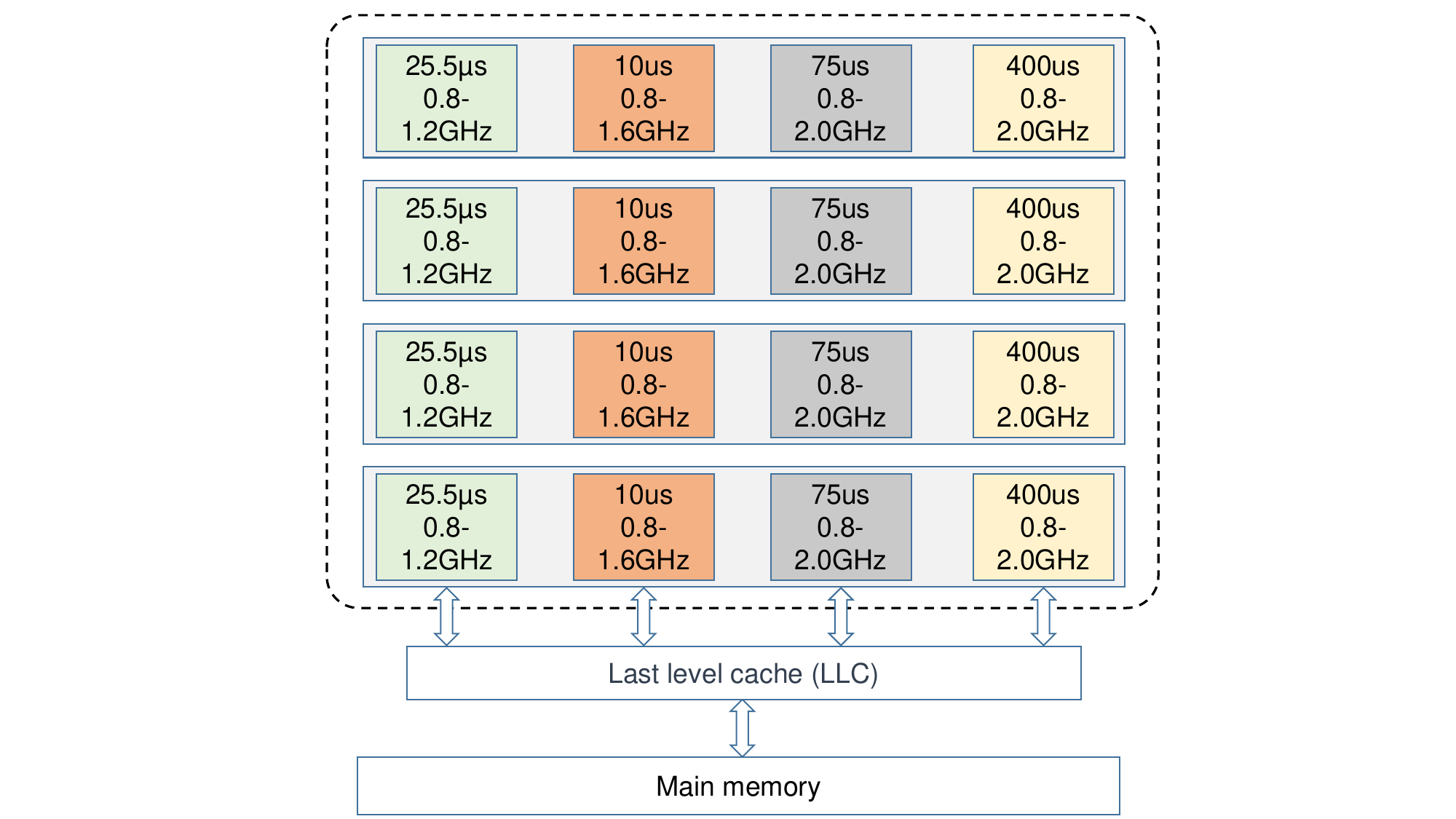}
      \vspace{-3pt}
      \caption{Scaling ARC for 16 cores featuring clusters of four cores. Each core has a different frequency range and cache retention time to satisfy different application requirements}
      \vspace{-5pt}
      \label{fig:scalability}
\end{figure}

In this paper, we do not provide a comprehensive study of ARC's scalability to more than four cores; we leave this study for future work. However, in this subsection, we briefly describe how we anticipate ARC to scale beyond four cores to many-core systems.

There are multiple alternatives for scaling ARC beyond four cores, depending on the specific use case. Figure \ref{fig:scalability} illustrates a candidate design for a sixteen-core system. Using the analysis described in the Section \ref{base_arc}, we empirically found that a set of four cores could optimize a variety of applications. We anticipate that for most systems, even in many-core systems, a subset of configurations will suffice to optimize the energy and/or performance for the variety of applications that execute on such systems. As such, ARC can be scaled to more cores simply by replicating the design in clusters of cores. In Figure \ref{fig:scalability}, for example, each core's parameters are similar to the base ARC architecture and all the four different cores within each cluster are similarly configured. The advantage of this approach is that the prediction technique for the base ARC architecture can be applied to this architecture as well. In a system where a different set of configurations are required, as dictated by the executing applications or application domains, ARC's prediction model may need to be updated to support the new labels. We do not anticipate that ARC will introduce substantial communication traffic; thus, the system's interconnection network can also be used for ARC's traffic without creating a bottleneck in the system. 

\begin{figure*}[t]
      \centering
      \includegraphics[width=0.8\linewidth]{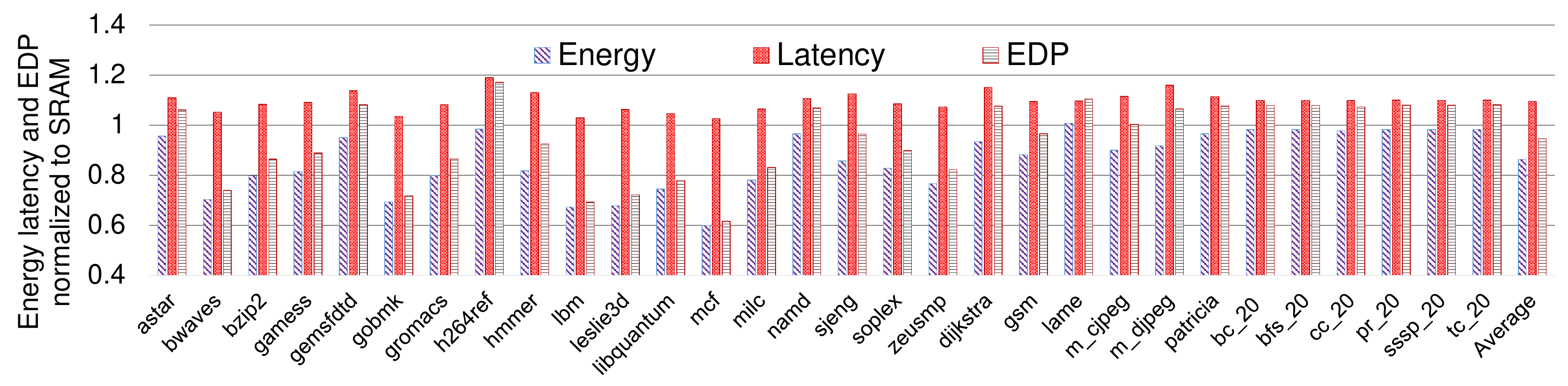}
      \vspace{-3pt}
      \caption{Energy savings potential of ARC vs a processor with SRAM caches (baseline of one) for different benchmarks. Results are with respect to the whole processor}
      \vspace{-5pt}
      \label{fig:application_ana}
\end{figure*}

\section{Experimental Setup} \label{sec:setup}

For our simulations, we modeled a quad-core ARM processor with parameters similar to Cortex A-53 using an in-house modified version of the GEM5 simulator \cite{gem5}. The modified GEM5 models the behavior of relaxed retention STT-RAM L1 caches with specified retention times. Each core has private 32KB, 4-way L1 instruction and data caches, with 64B line sizes. The data cache retention times used were: 10$\mu$s, 26.5$\mu$s, 75$\mu$s and 400$\mu$s, while 100$ms$ was used for all instruction caches. Through extensive design space exploration, we found that these retention times sufficed for the average cache block lifetimes of the benchmarks used in our experiments. The core frequencies ranged from 0.8 GHz to 2 GHz with an incremental step size of 0.2 GHz, and voltages ranged from 0.9V to 1.35V similar to prior work \cite{dvfs_volt_range}. The specific asymmetry of the STT-RAM cache retention times, frequency ranges, and cache write access cycles featured in the modeled cores are shown in Figure \ref{fig:arch}. We modeled unit cache read cycles for all STT-RAM caches. For comparison to SRAM, we assumed unit cache read and write access cycles.

To model the STT-RAM and SRAM cache energy, we used NVSim \cite{NVSim} integrated with the GEM5 statistics. The NVSim and GEM5 statistics were then integrated with the McPAT \cite{mcpat} simulator to model the energy consumption of the whole processor. Additional details of the modeled processor and cache configurations are shown in Table \ref{tab:retention}.

To represent a variety of workloads, we used 30 benchmarks from three benchmark suites: 18 from SPEC 2006\footnote{We were unable to use the full suite due to compilation errors.} \cite{spec2006}, 6 from MiBench \cite{mibench} and 6 benchmarks from GAP \cite{beamer2015gap}. We used thirteen arbitrary SPEC benchmarks for training, and the rest, combined with MiBench and GAP, for runtime experiments. Each benchmark was run for a maximum of 1 billiion instructions (some benchmarks, e.g., from MiBench have fewer than 1 billion instructions). To model different performance constraints, we assumed that the strictest constraint could be met using the fastest core and relaxed the constraints by slacks of 10\% and 20\%. Performance constraints can be determined by designer-specified deadlines or implicitly specified by the executing application \cite{hughes01}. We used the Scikit learn (Sklearn) \cite{scikit-learn} library in python to implement the decision trees in our model.

\section{ARC Evaluation} \label{sec:evaluation}
We evaluate the proposed ARC architecture by comparing with SRAM, a processor with homogeneous STT-RAM caches, and prior work that proposed logically adaptable retention STT-RAM caches (LARS) \cite{LARS8342053}. We implemented LARS in a multicore scenario for comparsion to our work. Note that the results presented herein are with respect to the \textit{whole} processor, which, as expected, are substantially less savings than cache-specific evaluations. To enable per-application evaluation, we assume a system with preemption; thus, applications can preempt background tasks in order to run on the predicted core. This is a realistic assumption for common resource-constrained devices, such as smartphones \cite{hahn17}. Section \ref{sec:multiprogrammed} evaluates ARC in the context of multiprogrammed workloads.

\subsection{Comparison to SRAM}

First, we evaluate the \textit{energy savings potential} of the ARC architecture by comparing with a processor featuring SRAM caches. Figure \ref{fig:application_ana} depicts the energy and latency achieved using the ARC architecture  normalized to SRAM. On average across all the benchmarks, ARC reduced the energy by 13.66\% compared to SRAM, with energy savings as high as 40.02\% for $mcf$. The average cache-only energy savings was 39.21\%. We observed that ARC performed best for workloads that exhibited low write accesses and high miss rates. As a result, ARC generally achieved higher energy savings for the SPEC and MiBench benchmarks than for the GAP benchmarks. However, ARC outperformed SRAM for all the benchmarks with respect to energy. Compared to SRAM, ARC improved the processor's energy-delay product (EDP) by an average of 5.44\% with savings as high as 38.45\% for $mcf$. Considering just the cache, ARC reduced the average energy by 39.21\% compared to a baseline SRAM cache, with energy optimization as high as 68.90\% for $mcf$ (detailed graphs omitted for brevity).

The energy savings was achieved at the cost of some latency overhead. Compared to SRAM, ARC degraded the latency by 9.52\% on average across all the benchmarks, with degradations as high as 18.95\% for $h264ref$. SRAM achieved these latency benefits because of lower write access cycles and lower cache miss rates. However, we note that the degradation with respect to SRAM is not a unique flaw of our proposed approach. Prior work observed similar trends. For instance, Cheng et al. \cite{sram_performance_degrade} observed 24\% degradation in memory access latency resulting from replacing SRAM with STT-RAM, prompting their proposal to use a hybrid (SRAM + STT-RAM) cache to reduce the performance overhead. The hybrid cache is orthogonal to our work, and the latency overheads can be reduced by augmenting the cache asymmetry proposed herein by using SRAM in some of the cores.  


\subsection{{Comparison to Homogeneous Retention STT-RAM and Exhaustive Search}\label{section:deadline_resutls}}

\begin{figure*}[t]
        \centering
        \begin{subfigure}[b]{0.475\textwidth}   
            \centering 
            \includegraphics[width=0.85\textwidth]{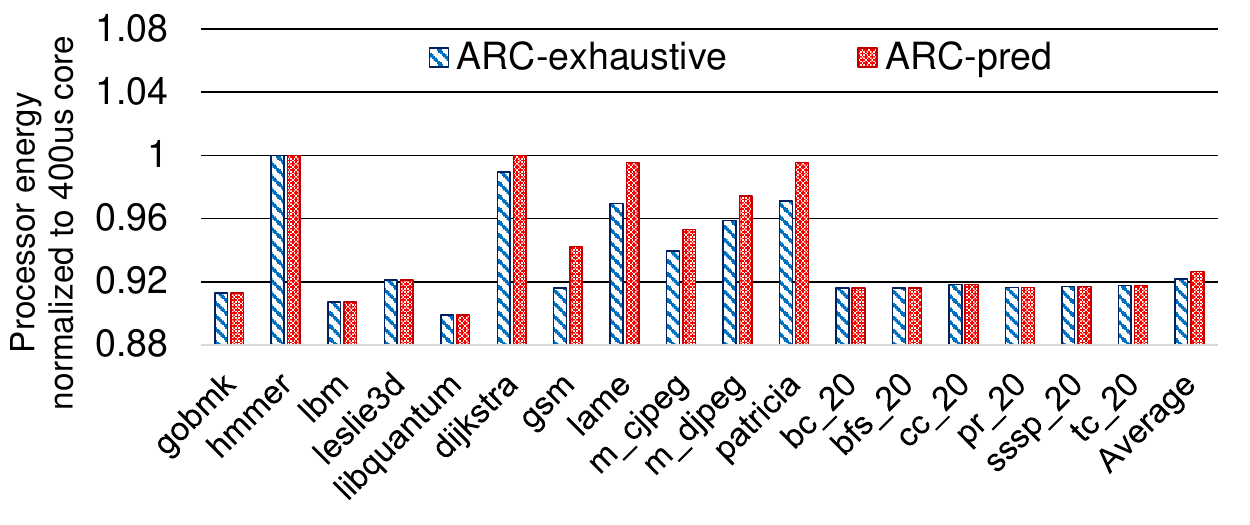}
            \caption{No constraint}
             
            \label{fig:no_deadline}
        \end{subfigure}
        \hfill
        \begin{subfigure}[b]{0.475\textwidth}   
            \centering 
            \includegraphics[width=0.85\textwidth]{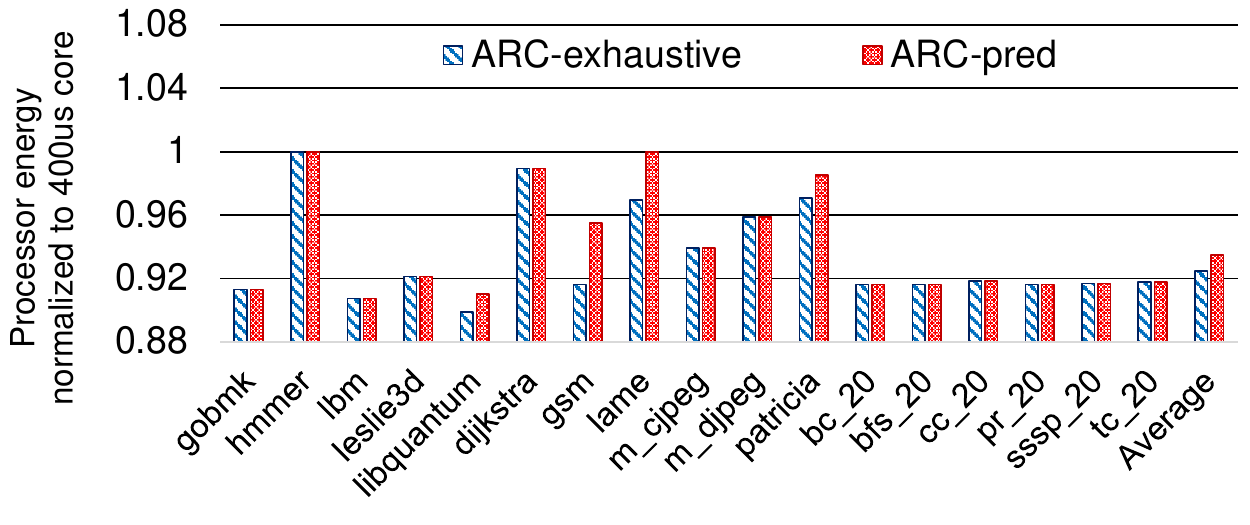}
            \caption{20\% slack}%
               
            \label{fig:deadline20}
        \end{subfigure}
        \vskip\baselineskip
        \begin{subfigure}[b]{0.475\textwidth}  
            \centering 
            \includegraphics[width=0.85\textwidth]{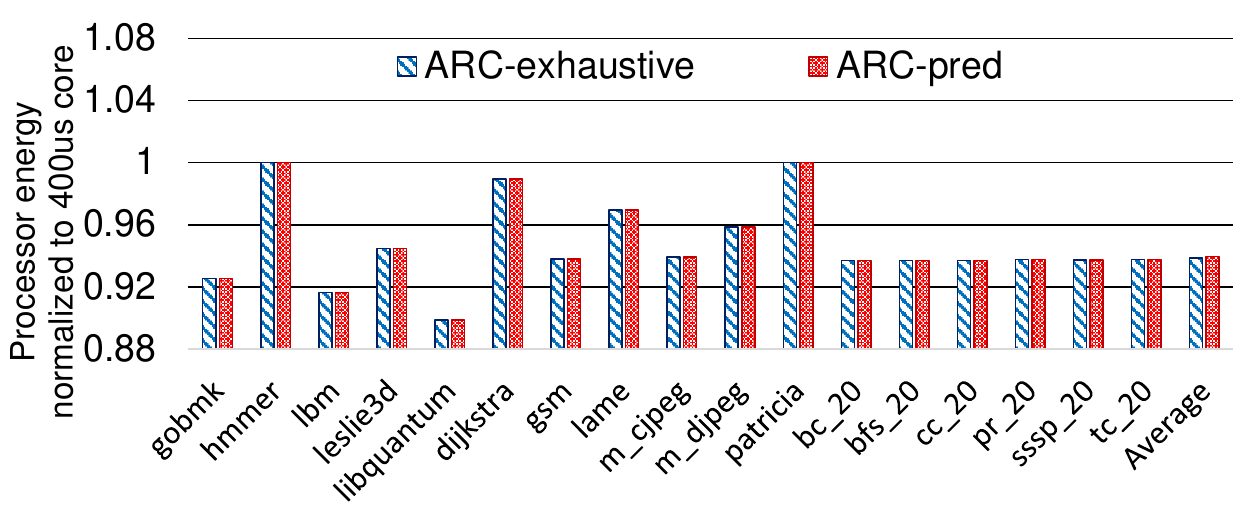}
            \caption{10\% slack}
               
            \label{fig:deadline1}
        \end{subfigure}
        \quad
        \begin{subfigure}[b]{0.475\textwidth}
            \centering
            \includegraphics[width=0.85\textwidth]{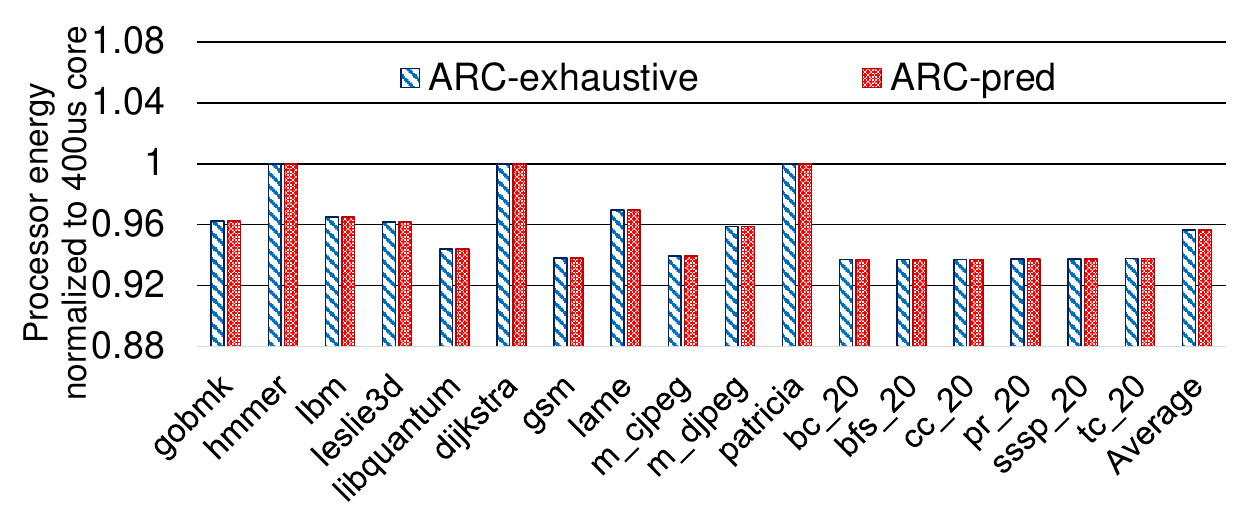}
            \caption{Best performance required}
            \label{fig:strict_Deadline}
        \end{subfigure}
        \caption{ARC energy savings normalized to a base processor featuring the 400$\mu$s STT-RAM cache on all cores (baseline of one) for different performance constraints. Results are with respect to the whole processor}
        
        \label{fig:deadline_results_main}
    \end{figure*}

In this section, we compare the energy savings of the ARC prediction model to a processor with an optimized homogeneous retention time and to a system called \textit{ARC-exhaustive}. In the homogeneous system, the best average retention time is determined via a priori design space exploration and featured in all the cores. In ARC-exhaustive, sampling is used to determine the \textit{best} core for each application. For rigorous experimentation, our training data comprised of 13 benchmarks from SPEC benchmark suite, and our test data comprises of 17 benchmarks: five from SPEC, six from MiBench, and six from GAP. We evaluated the energy savings in the context of four performance constraints: no constraint, 20\% slack, 10\% slack, and best performance required. The results obtained by our ARC prediction model are compared with both the homogeneous system and with ARC-exhaustive.

Figure \ref{fig:deadline_results_main} presents the energy savings of exhaustive search (ARC-exhaustive) and the prediction model (ARC-pred) normalized to a processor featuring 400$\mu$s (Core 4 in Figure \ref{fig:arch}) STT-RAM caches across all the cores. We used the 400$\mu$s for comparison, since it achieved the lowest miss rates on average across all the benchmarks. On average, ARC-pred achieved energy savings of 7.35\%, 6.51\%, 6.04\% and 4.34\% for no constraint, 20\% slack, 10\% slack, and best performance, respectively, and the results were within 0.57\% of ARC-exhaustive, on average. Energy savings were up to 10.11\% for the 10\% slack for $libquantum$. We reiterate that the comparisons are with respect to the whole processor, and the energy savings are a lot higher (> 20\% on average) when considering just the cache. 

Whereas ARC-pred performed similar to ARC-exhaustive for most benchmarks, there were a few benchmarks where ARC-pred degraded the energy compared to ARC-exhaustive. In the worst case, ARC-pred degrade the energy by up to 3.04\% for $lame$ under 20\% slack, due to false prediction; however, our prediction approach never degraded the energy savings compared to the base core as a result of the feedback process (Section \ref{sec:approach}). 

As expected, ARC's energy savings decrease as the performance constraints become more stringent. As seen in Figures \ref{fig:no_deadline}, \ref{fig:deadline20}, and \ref{fig:deadline1}, as the performance constraint becomes stricter, the energy savings also decrease. Overall, ARC-pred met the constraints for all the benchmarks (Figures \ref{fig:deadline1} and \ref{fig:deadline20}). In a few cases (e.g., for \textit{patricia} with 10\% slack), constraints were initially violated due to mispredictions. However, ARC was able to correct the best core during the feedback process. 

We note that the processor energy savings achieved is a result of the proper mapping of applications to the respective cores. This mapping enabled right-provisioned clock frequencies in consonance with an optimal number of write cycles among the available cores. Notably, we also explored using ARC for cache-specific savings and found that ARC reduced the cache energy by more than 20\% in the different execution scenarios. However, we also found that the best configurations for cache-specific energy optimization were not necessarily the best when considering the whole processor. As such, we focused our analysis on the processor-wide results. 

\subsection{ARC Evaluation in a Multiprogrammed Scenario} \label{sec:multiprogrammed}

\begin{table}[t]
\vspace{-10pt}
\renewcommand{\arraystretch}{0.65}
\caption{Multiprogrammed workload distribution}
\label{tab:multi_label}
\centering
\vspace{-5pt}
\scalebox{0.75}{
\begin{tabular}{|c|c|c|c|c|c|c|}
    \hline
    \# & Workload1 &Workload2 &Workload3 &Workload4 &Workload5 &Workload6\\  
    \hline
    1 & dijkstra	&lbm	&gsm	&cc\_20	&patricia	&hmmer\\
    2 & pr\_20	&hmmer	&gobmk	&bc\_20	&cc\_20	&m\_cjpeg\\
    3 & m\_djpeg	&m\_cjpeg	&sssp\_20	&leslie3d	&pr\_20	&tc\_20\\
    4 & lame	&bfs\_20	&libquantum	&tc\_20	&libquantum	&bc\_20\\
    \hline
\end{tabular}}
\vspace{-2pt}

\end{table}

\begin{figure}[t]
    \vspace{-10pt}
      \centering
      \includegraphics[width=0.8\linewidth]{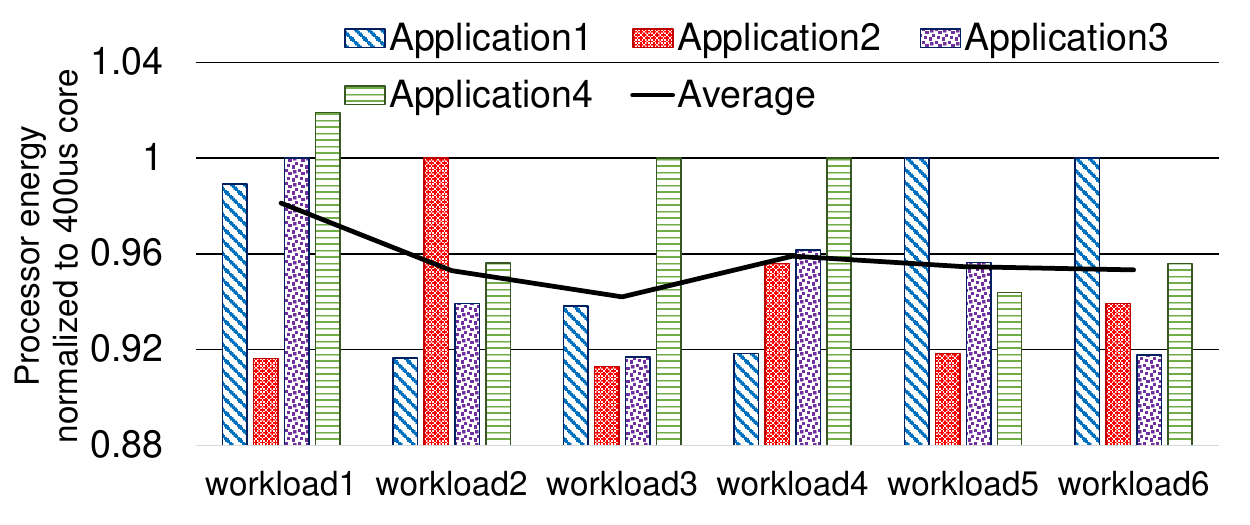}
      \vspace{-3pt}
      \caption{ARC energy savings in a multi-programmed scenario. We present results assuming no performance constraints are specified. If the predicted core is occupied, the next best available core is selected}
      \label{fig:multi_label}
      
\end{figure}

In this section, we evaluate ARC in a multiprogrammed execution scenario. We created six multiprogrammed workloads comprising of four randomly selected benchmarks from our 17 test applications, ensuring that all 17 benchmarks are represented in the workloads. Table \ref{tab:multi_label} depicts the workload compositions used. Unlike for previous experiments, we assume no preemption in this case. That is, if the predicted best core is unavailable, the next best available core is used. As such, we augmented the classifier in our prediction model to include a ranking of the output labels. 

Figure \ref{fig:multi_label} depicts the overall processor energy of ARC normalized to a homogeneous STT-RAM design. The average energy savings was 1.88\%, 4.70\%, 5.80\%, 4.10\%, 4.53\% and 4.67\% for workloads one through six, respectively. The average energy savings across all the workloads was 4.28\%. Overall, energy savings were lower since some benchmarks were forced to run on sub-optimal cores. We also observed higher energy consumption than the base core for $workload1$ due to the unavailability of the best core for application 4 ($lame$). While these results illustrate the energy savings potentials of ARC, we note that some nuances are ignored herein. For instance, the impact of preemption can be studied further; i.e., benefits of waiting on the best core vs. running on a sub-optimal core. These nuances become more important in performance-constrained scenarios. We plan to explore these nuances in future work.

\subsection{Comparison to Prior Work}
We compared ARC-pred to a multicore implementation of the logically adaptable retention STT-RAM (LARS) cache proposed in \cite{LARS8342053}. LARS features four STT-RAM units with different retention times within a single chip to enable specialization to applications' needs. We observed similar energy savings between ARC and LARS (graphs omitted for brevity). However, notably, ARC only features a single STT-RAM unit per core, thereby reducing the per-core physical L1 cache area by 75\% compared to LARS. Furthermore, we compared our work to prior art with respect to the migration overheads. Each migration in LARS took approximately 10$\mu$s. By directly predicting the best core, ARC reduced the migration overheads by 67.80\%, on average, compared to LARS. 

\subsection{ARC Overheads}

For brevity, we only report overheads for when no performance constraints are specified, since the performance-constrained scenarios are extensions of this. To minimize hardware overheads, we assume a software implementation of the prediction model and the history table (Section \ref{sec:approach}). For both the model and history table (assuming a 120-entry table), ARC consumed a total of 0.0167MB of RAM. The history table was large enough for all the applications considered in our experiments, and the overhead included the models for the different performance constraints. When the history table is full, however, entries can be replaced using a replacement policy, such as least recently used. As described in Section \ref{base_arc}, the monitor counter introduced a 2-bit per block overhead, resulting in a total overhead of 128 B per core (i.e., a two-block overhead per core). 

We also evaluated ARC overheads with respect to the prediction time. On average across all the benchmarks, the prediction time---including collecting the hardware characteristics and running the ARC prediction algorithm---was 3.23$\mu$s. We analyzed the migration overheads and found that, on average, ARC accrued migration overheads of 7.94$\mu$s. We plan to explore the tradeoffs of a hardware implementation in future work.

\section{Conclusion and Future Work}
In this paper, we explored the behavior of STT-RAMs with variable clock frequencies, with respect to applications' runtime cache block requirements. Our analysis showed that, unlike SRAM caches, STT-RAMs exhibit substantial variability regarding the best clock frequencies; the applications' retention time requirements must also be taken into consideration. Thus, to enable energy savings by exploiting the interplay of DVFS and variable retention time requirements, we proposed the \textit{Asymmetric-Retention Core (ARC)} architecture for multicore systems. ARC features STT-RAM caches with different retention times, clock frequencies, and access cycles in different cores, such that applications are run on the core that best satisfies the application's needs. We also proposed a runtime decision tree-based ARC prediction model that directly predicts the best core on which to run an application, based on the application's execution characteristics. Using extensive simulations and experiments with several execution scenarios, results reveal that ARC can reduce the average cache energy by 39.21\% and the overall processor energy by 13.66\%, compared to a system with SRAM caches. Compared to a system with optimized homogeneous STT-RAM cache, ARC can reduce the average cache energy and overall processor energy by 20.19\% and 7.66\%, respectively. Future work includes exploring ARC in more complex many-core heterogeneous systems with multilevel cache hierarchies. 

\section*{Acknowledgement}
This work was supported in part by the National Science Foundation under grant CNS-1844952 (CAREER). Any opinions, findings, and conclusions or recommendations expressed in this material are those of the authors and do
not necessarily reflect the views of the National Science Foundation.
	\balance
	\bibliographystyle{ACM-Reference-Format}
	\bibliography{refs}

\end{document}